\documentclass[showpacs,twocolumn,bibnotes,footinbib]{revtex4}

\usepackage{graphicx}
\usepackage{amsmath}

\begin{document}

\title{Mach-Zehnder-Fano interferometer}

\author{Andrey E. Miroshnichenko}
\affiliation{Nonlinear Physics Center and Center for Ultra-high bandwidth Devices for Optical Systems (CUDOS),
Research School of Physics and Engineering, Australian National University, Canberra ACT 0200, Australia}

\author{Yuri S. Kivshar}
\affiliation{Nonlinear Physics Center and Center for Ultra-high bandwidth Devices for Optical Systems (CUDOS), Research School of Physics and Engineering, Australian National University, Canberra ACT 0200, Australia}

\begin{abstract}
We introduce a concept of the Mach-Zehnder-Fano interferometer by inserting a cavity exhibiting Fano resonance into a conventional interferometer. By employing the scattering-matrix approach, we demonstrate that the transmission is sensitive to a position of the cavity such that an asymmetric structure exhibits a series of narrow resonances with almost perfect reflection. We discuss how to implement this novel geometry in two-dimensional photonic crystals and use direct numerical simulations to demonstrate novel regimes of the resonant transmission and reflection.
\end{abstract}


\maketitle

The Mach-Zehnder interferometer (MZI) was suggested more than a hundred years ago, and it is still a key element in integrated optics being implemented into design of various photonic devices including modulators, switches, filters, and multiplexers. The recent developments moved this concept toward nanoscales, and several studies employed ring-resonator structures in order to low the power threshold~\cite{zhou,hitoshi,rostami,inoue,liu,cho,fathi,shih} which can demonstrate very efficient nonlinear response and can be easily tuned. A novel advantage of using additional resonant cavities has been also discussed~\cite{yljyxlpw:OL:05,lzawp:OL:07,landobasa1,landobasa2,li}.

In this Letter we introduce a concept of the Mach-Zehnder-Fano interferometer that can be realized on various platforms such as photonic crystals or ring resonators that possess effective discreteness of periodic photonic structures for engineering the transmission properties. Side-coupled cavity with a resonant state in the transmission band provides a sharp variation of the scattering phase, which allows to alter the transmission properties of the whole structure in a narrow frequency range. Photonic crystals allow to implement this idea in the form of very compact structures.

The physics of the Fano resonance is explained by an interference between a continuum and discrete state~\cite{fano}. The simplest realization is a one-dimensional discrete array (continuum state) with a side-coupled defect (discrete state). In such a system scattering waves can either bypass the defect or interact with it~\cite{review}. The $\pi$ phase shift at the resonance results in destructive interference at the output, leading to resonant suppression of the transmission. Usually, each Fano resonance can be associated with a particular resonant state of the side-coupled defect~\cite{mir2}. Similarly, we demonstrate below that MZI with a symmetric Fano defect displays a resonant state with a well-defined suppression of the transmission. However, for an asymmetric Fano defect the transmission exhibits more than one resonance, and it can be understood in terms of the interaction between two continua with one discrete state~\cite{fano}. Indeed, both arms of MZI can be considered as two continua, and the side-coupled defect as a discrete state. Since both the arms are coupled, the defect becomes effectively coupled to both of them but with different strengths. As a result, we may obtain two resonances instead of one.  We introduce the main concept for an effective discrete model, and then apply it to photonic-crystal based devises whose optical properties can be modeled by an effective discrete model derived by the Green's function formalism~\cite{mingaleev}.

\begin{figure}
\includegraphics[width=0.9\columnwidth]{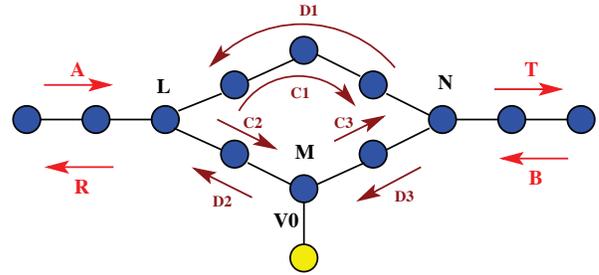}%
\caption{\label{fig:fig1} (Color online)
Schematic structure of the Mach-Zehnder-Fano interferometer described by an effective discrete model. $A$ and $B$ are incoming, $T$ and $R$ are scattered amplitudes. $C_i$ and $D_i$ are amplitudes of the forward and backward propagating waves. $V_0$ is the coupling to Fano defect. $L$, $N$, and $M$ stand for the number of sites.}
\end{figure}

First, we study the transmission properties of the MZI with a side-coupled Fano defect in a generic discrete model, as shown schematically in  Fig.~\ref{fig:fig1}. We assume the nearest-neighbor interaction between the discrete sites. In all parts of the structure the propagating waves can be separated onto forward and backward ones, and characterized by the corresponding amplitudes. The relation between the amplitude of the incoming and outgoing waves can be easily written in terms of the scattering-matrix approach,
\begin{eqnarray}
\label{eq:eq1}
\left(
  \begin{array}{c}
  R\\C_1\\C_2
  \end{array}
\right)
  ={\mathbf S}^Y_L
  \left(
  \begin{array}{c}
  A\\D_1\\D_2
  \end{array}
\right)\;,\;
\left(
  \begin{array}{c}
  T\\D_1\\D_3
  \end{array}
\right)
  ={\mathbf S}^Y_{-N}
  \left(
  \begin{array}{c}
  B\\C_1\\C_3
  \end{array}
\right)
\end{eqnarray}
with the scattering matrix of the $Y$-splitter of the form
\begin{eqnarray}
\label{eq:eq2}
{\mathbf S}^Y_K=\left(
  \begin{array}{ccc}
  e^{-2iKq}r_Y&t_Y&t_Y\\
  t_Y&e^{2iKq}r_Y&e^{2iKq}t_Y\\
  t_Y&e^{2iKq}t_Y&e^{2iKq}r_Y
  \end{array}
\right)\;,
\end{eqnarray}
where $K=L,-N$, while $t_Y=2/(3-i\cot q)$ and $r_Y=t_Y-1$ are the transmission and reflection amplitudes of the $Y$-splitter~\cite{mir1}.

The relation between the scattering amplitudes of the lower arm with a side-coupled defect can be obtained via the transfer matrix
\begin{eqnarray}
\label{eq:eq3}
\left(
  \begin{array}{c}
  C_2\\D_2
  \end{array}
\right)
  ={\mathbf T}_M
  \left(
  \begin{array}{c}
  C_3\\D_3
  \end{array}
\right), \nonumber\\
{\mathbf T}_M=
\left(
  \begin{array}{cl}
  \omega_q-\frac{V_0^2}{\omega_q-E_F}&-1\\1&0
  \end{array}
\right),
\end{eqnarray}
where $\omega_q=2\cos q$, $q$ is wavenumber, $V_0$ is the coupling parameter of the side-coupled defect, and $E_F$ its eigenfrequency~\cite{mir2}.
By setting $A=1$ and $B=0$ one can  find numerically the transmission ($T$) and reflection ($R$) amplitudes by solving simultaneously Eq.~(\ref{eq:eq1}) and Eq.~(\ref{eq:eq3}).

We consider the conventional MZI with the two symmetric arms of the same length. In the absence of a side-coupled defect (i.e. $V_0=0$),
the transmission coefficient exhibits a periodic dependence, in an analogy with the Fabry-Perot resonance with zero phase difference at the output of both arms $\delta\phi=\phi_1-\phi_2\equiv0$, where $\phi_1$ and $\phi_2$ are the phases of the  lower and upper arms of MZI, correspondingly. The number of peaks is proportional to the length of both the arms. In the geometry under consideration the phase accumulation is linearly proportional to the length of the arms $\phi_{1,2}\propto q(N-L)$. The purpose of our study is to demonstrate that by adding a side-coupled defect we may alter quite substantially this dependence due to presence of the Fano resonances. The main feature of the Fano resonance is  suppression of the transmission. Side-coupled defect provides with an additional resonant path. The scattering waves accumulate $\pi-$phase shift in the vicinity of a resonance.  As a result there is a sharp $\pi$ jump of phase difference of  scattered waves  at the resonance, which lead to destructive interference and resonant suppression of the transmission. This strong dependence of the phase in the vicinity of the resonance will allow one to design the MZI with shorter arms, making the final device much more compact.

\begin{figure}
\includegraphics[width=1.\columnwidth]{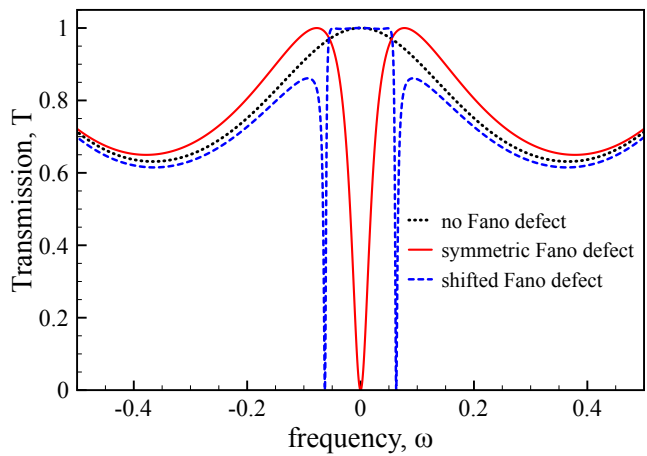}%
\caption{\label{fig:fig2} (Color online)
Transmission coefficient of the discrete MZI model for: conventional interferometer (dotted) and the interferometer interacting with a cavity for
symmetrically (solid) $M=5$ and asymmetrically (dashed) $M=4$ placed defect. Other parameters are: $L=1$, $N=11$, $V_0=0.2$, and $E_F=0$. }
\end{figure}

When a cavity is coupled to the straight waveguide, the Fano resonant reflection ($T=0$) is observed at the eigenfrequency of the cavity mode, $\omega_q=E_F$. In the case of the MZI, it is no longer true. Indeed, by attaching the Fano defect symmetrically to the MZI the resonance still takes place at the eigenfrequency of the defect $E_F$. In this spatially symmetric case the coupling to the upper arm is somehow reduced, and there is only one resonance of the Fano defect with the lower arm. But, the situation is drastically changed, when the Fano defect is attached asymmetrically (see Fig.\ref{fig:fig2}). In this situation there are two Fano resonances. Each resonance can be associated with the interaction of the side-coupled defect with one of MZI arms. When the defect is placed symmetrically the waves propagating clockwise and counterclockwise acquire  the same phase accumulation, and, eventually, canceling each other leading to zero effective coupling with the upper arm. That's why the resonance takes place at the eigenfrequency of the defect, in the similarity to the straight waveguide. For asymmetrically placed Fano defect clockwise and counterclockwise waves acquire different phase accumulation resulting in different resonant conditions. The side-coupled defect can be considered as an additional degree of freedom for scattered waves to propagate, making  the MZI to consist of three arms, instead of two. It turns out that the role of this effective arm is significant for the final outcome due to sharp variation of the phase.

\begin{figure}
\includegraphics[width=1.\columnwidth]{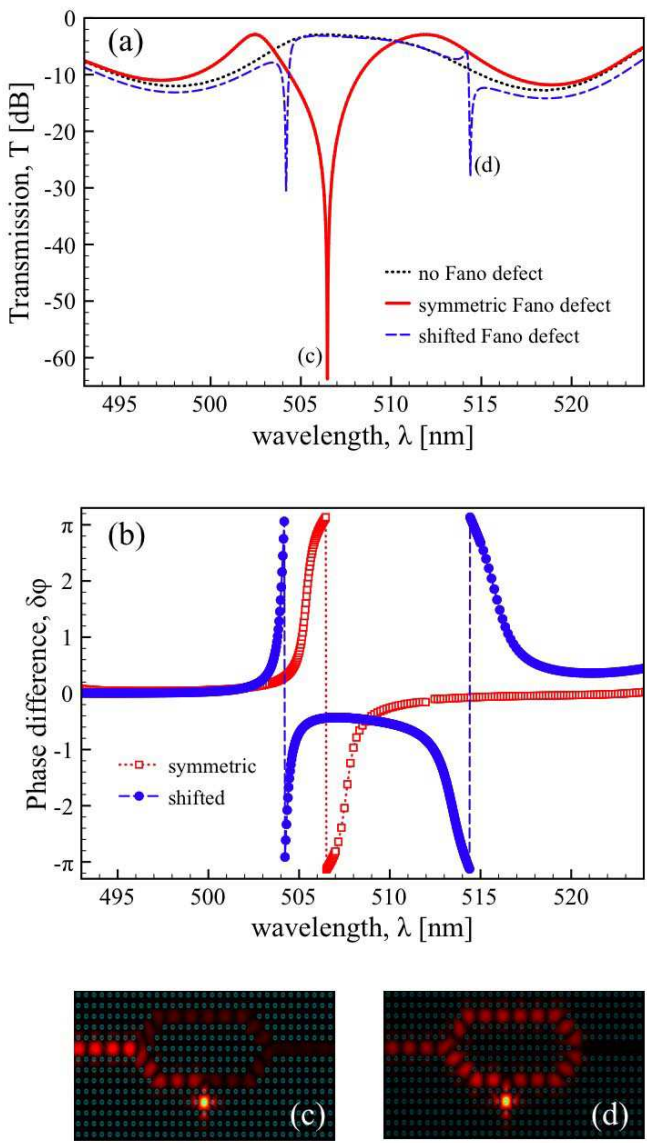}%
\caption{\label{fig:fig3} (Color online)
Direct numerical simulations of the Mach-Zehnder-Fano interferometer realized in a two-dimensional photonic crystal. (a)
transmission of the conventional MZI (dotted), and the MZI with a cavity paced either symmetrically (solid) or asymmetrically (dashed)
(cf. Fig.~\ref{fig:fig2}). (b) Wavelength dependence of the phase difference at the output of two MZI arms.
The plots (c) and (d) demonstrate the electric field distribution $|E_z|$ at various resonances marked in plot (a).}
\end{figure}

As one of the possible realizations of this concept, we consider a two-dimensional photonic crystal created by dielectric rods in air with radius $r=35$~nm and the refractive index $n_r=3.14$, arranged in a square periodic lattice of the radius $a=187.5$~nm in $(x,y)$ plane. Left and right waveguides, and both arms of the MZI are made by removing certain rods, as depicted in Fig.~\ref{fig:fig3}. For given parameters, the photonic crystal waveguide supports TE waves with the frequencies from the visible range. The side-coupled defect is made by replacing one the dielectric rod with a polymer rod of the same radius and refractive index $n_d=1.63$. Such kind of defect supports a quadrupole mode in the visible range at $\lambda_{\rm quad}\approx507$~nm.

The conventional Mach-Zehnder structure exhibits the periodic dependence of the transmission coefficient vs. the wavelength. In the current realization the maximum of the transmission is limited by 3dB due to $Y$-splitter limitations [see Fig.~\ref{fig:fig3}(a)]. By adding a side-coupled defect symmetrically to one of the arms of MZI, we make that the transmission coefficient drops resonantly near the quadrupole mode of the defect $T(\lambda_{\rm quad})\rightarrow0$, in accordance with above considerations. Far away from the resonance the transmission curve resembles the one for the pure MZI without the defect. By shifting the position of the side-coupled defect by one site to the left, the transmission coefficient drastically changes. Instead of one Fano resonance there are two of them, which are much narrower than in the symmetric case. The difference between two resonances is revealed by looking at the  electric field distribution $|E_z|$. In the symmetric case the electric field is mostly concentrated at the lower arm [see Fig.~\ref{fig:fig3}(c)], which indicates that only one arm is involved in the interaction at the resonance. On contrary, in the shifted case two arms are equally excited [see Fig.~\ref{fig:fig3}(d)], indicating that both of them are interacting with the side-coupled defect. This is in full agreement with theoretical results for discrete model considered above. It can be further confirmed by looking at the phases difference at the output of two arms $\delta\phi$. All resonances are accompanied by $\pi$ phase difference of scattering waves in both arms $|\delta\phi|\sim\pi$ [see Fig.~\ref{fig:fig3}(b)]. But what is important now is the sign of the phase difference. In the symmetric case the phase of the lower arm $\phi_1$ is increased by $\pi$ in the vicinity of the resonance, while the phase of the upper arm $\phi_2$ varies slowly. It is characterized by jump from $\pi$ to $-\pi$ of the phase difference. It again indicates the effective interaction between the side-coupled defect and the lower arm only.
In the shifted case in addition to this behavior there is opposite one when the phase of the upper arm $\phi_2$ acquires increase by $\pi$, which is characterized by jump from $-\pi$ to $\pi$ of the phase difference. Thus, in this case the side-coupled defect is effectively interacting with the upper arm than with the lower one.

In conclusion, we have discussed the properties of a novel Mach-Zehnder-Fano interferometer created by inserting a cavity into the conventional MZI.
We have demonstrated that the excitation of the cavity mode changes dramatically the transmission characteristics of the structure such that the coupling between the arms is effectively canceled due to the Fano-induced destructive interference. By changing the position of the cavity, the transmission coefficient can be altered dramatically. We have implemented this novel Mach-Zehnder-Fano geometry in two-dimensional photonic crystals, and observed a striking similarity between the transmission characteristics of the effective discrete model and direct numerical simulations.

The authors acknowledge a support from the Australian Research Council.

\end{document}